# Crystal structure of cobalt hydroxide carbonate $Co_2CO_3(OH)_2$: density functional theory and X-ray diffraction investigation

**Jorge González-López[a], Jeremy K. Cockcroft[b], Ángeles Fernández-González[a], Amalia Jimenez[a] and Ricardo Grau-Crespo[c]***

[a] Department of Geology, University of Oviedo, Calle Jesús Arias de Velasco, s/n, Oviedo, 33005, Spain

[b] Department of Chemistry, University College of London, 20 Gordon St., London, WC1H 0AJ, United Kingdom

[c] Department of Chemistry, University of Reading, Whitenights, Reading, RG6 6AD, United Kingdom

Correspondence email: r.grau-crespo@reading.ac.uk

**Synopsis** The crystal structure of cobalt carbonate hydroxide $Co_2CO_3(OH)_2$, a solid important in materials and environmental science, is investigated using Density Functional Theory (DFT) simulations and Powder X-Ray Diffraction (PXRD) measurements.

**Abstract** The cobalt carbonate hydroxide $Co_2CO_3(OH)_2$ is a technologically important solid which is used as precursor for the synthesis of cobalt oxides in a wide range of applications, and it also has relevance as a potential immobilizer of toxic element cobalt in the environment, but its detailed crystal structure is so far unknown. We have investigated the structure of $Co_2CO_3(OH)_2$ using Density Functional Theory (DFT) simulations as well as Powder X-Ray Diffraction (PXRD) measurements on samples synthesized via deposition from aqueous solution. We consider two possible monoclinic phases, with closely related but symmetrically different crystal structures, based on those of the minerals malachite ($Cu_2CO_3(OH)_2$) and rosasite ($Cu_{1.5}Zn_{0.5}CO_3(OH)_2$), as well as an orthorhombic phase that can be seen as a common parent structure for the two monoclinic phases, and a triclinic phase with the structure of the mineral kolwezite ($Cu_{1.34}Co_{0.66}CO_3(OH)_2$). Our DFT simulations predict that the rosasite-like and the malachite-like phases are two different local minima of the potential energy landscape for $Co_2CO_3(OH)_2$, and are practically degenerate in energy, while the orthorhombic and triclinic structures are unstable and experience barrierless transformations to the malachite phase upon relaxation. The best fit to the PXRD data is obtained using a rosasite model (monoclinic with space group $P112_1/n$ and cell parameters $a = 3.1408(4)$ Å, $b = 12.2914(17)$ Å, $c = 9.3311(16)$ Å, $\gamma = 82.299(16)°$). However, some features of the PXRD pattern are still not well accounted for by this refinement and the residual parameters are relatively poor. We discuss the relationship between the rosasite and malachite phases of $Co_2CO_3(OH)_2$ and show that they can be seen as polytypes. Based on the similar calculated stability of these two polytypes, we speculate that some level of stacking disorder could account for the poor fit of our PXRD data. The possibility that $Co_2CO_3(OH)_2$ could crystallize, under different growth conditions, as either rosasite or malachite, or even as a stacking-disordered phase intermediate between the two, requires further investigation.

1. Introduction

The solid structure of cobalt (II) carbonate hydroxide ($Co_2CO_3(OH)_2$) is important for technological and environmental reasons. It is commonly used as a precursor in the synthesis of cobalt oxides (Li *et al.*, 2006, Li *et al.*, 2012, Xie *et al.*, 2010, Xu & Zeng, 2003), which have a wide range of technological applications as catalysts in the petroleum industry, magnetic materials, semiconductors, chemical gas sensors, solar collectors, lithium-ion batteries, etc. (Ando *et al.*, 1997, Robert *et al.*, 2005, Tuti & Pepe, 2008, Wang *et al.*, 2008, Yuan *et al.*, 2003, Yang *et al.*, 2011). $Co_2CO_3(OH)_2$ has also been proposed as a potential immobilizer of cobalt in the environment (Katsikopoulos *et al.*, 2008). Cobalt is considered as a possible carcinogen by the International Agency for Research on Cancer (IARC, 1991). Moreover, some of its isotopes ($^{58}Co$ and $^{60}Co$) are radioactive, which makes them useful in nuclear applications, but also imply risks to human health. Although cobalt appears only as a trace element in the earth crust (Smith & Carson, 1981), it can be found more abundantly in soils and groundwater as a consequence of the extract process of Co-bearing minerals, and also as waste derived from industrial activities, e.g. building (alloy steel), use of cobalt-containing fertilisers, manufacture of pigments, batteries, etc. (ATSDR, 2004). Previous research has considered possible routes for cobalt immobilization (*via* precipitation and/or interaction) by carbonate-containing materials, in particular calcite $CaCO_3$ (Katsikopoulos *et al.*, 2008, Wada *et al.*, 1995, Braybrook *et al.*, 2002). However, no clear incorporation in calcite has been observed. In fact, a theoretical study of the thermodynamic properties of $Ca_{1-x}Co_xCO_3$ solid solutions concluded that no significant amount of cobalt can be expected to incorporate substitutionally in the calcite structure under ambient conditions (González-López *et al.*, 2014). Since cobalt immobilization in aqueous environments via calcite precipitation seems to be difficult to achieve, there is interest in investigating other phases that could immobilize cobalt. The first substance precipitated from cobalt and carbonate ions in aqueous solution at ambient temperature is known to be an amorphous phase (Barber *et al.*, 1975). Katsikopoulos et al. (2008) reported that this amorphous substance corresponds to a hydrated cobalt carbonate. These authors showed that the precipitation from $Co^{2+}$ and $CO_3^{2-}$ at room temperature from aqueous solution, leads to the transformation from the amorphous carbonate to a carbonate phase with better crystallinity, through aging on the same aqueous solution from where it has been precipitated. Thus, amorphous and crystalline cobalt hydroxide carbonate phases are likely to exist in areas of the earth crust where Co is anomalously present (e.g. mining, waste disposal sites, etc.) in contact with ground and fresh waters, and might play an important role in cobalt immobilization in the environment.

The detailed crystal structure of $Co_2CO_3(OH)_2$ is so far unknown. A preliminary PXRD study by Wang *et al.* (2009) suggested a malachite-type monoclinic structure with space group $P12_1/a1$ and *a*=9.448 Å, *b*=12.186 Å, *c*=3.188 Å, β=91.879° but atomic positions were not refined. In a short conference report later (Wang *et al.*, 2010), these authors described a refinement attempt, but the reported positions are not correct (there are no defined $CO_3$ units nor $CoO_6$ octahedra), and are not comparable with those in the malachite structure. On the other hand, some of us have recently reported the PXRD characterization of synthetic $Co_2CO_3(OH)_2$ and indexed the structure as a rosasite-like monoclinic structure with space group $P12_1/a1$ and $a$=12.886 Å, $b$=9.346 Å, $c$=3.156 Å, $\beta$=110.358°, but we did not attempt to refine atomic positions either, due to the low crystallinity of the samples (González-López *et al.*, 2016).

As will be seen in more detail below, the malachite-like and rosasite-like structures, while closely related and expressed in the same space group, are not isotypic. The relationship between them has been discussed before by Girgsdies & Behrens (2012), where an orthorhombic structure with space group Pbam was also proposed as a common hypothetical parent structure (aristotype). Interestingly, some authors have assigned the $Co_2CO_3(OH)_2$ structure to the orthorhombic crystal system, although again no atomic positions were reported (Yang *et al.*, 2011, Xing *et al.*, 2008). Finally, there is also a triclinic structure associated with the $MCO_3(OH)_2$ stoichiometry, which is that of the mineral kolwezite ($Cu_{1.34}Co_{0.66}CO_3(OH)_2$), where the three cell angles are close to 90° (Deliens & Piret, 1980).

The objective of the present work is to elucidate the crystal structure of $Co_2CO_3(OH)_2$ using a combination of density functional theory (DFT) calculations and PXRD measurements on hydrothermally synthesized samples. We have investigated the thermodynamic stability of $Co_2(OH)_2CO_3$ in each of the two monoclinic phases (rosasite and malachite), in the orthorhombic aristotype structure, and in the triclinic kolwezite structure. We then use the DFT models to aid the interpretation of the PXRD patterns.

## 2. Methodology

### 2.1. Density functional theory calculations

The equilibrium geometries and energies of different possible phases of $Co_2CO_3(OH)_2$ were calculated using DFT simulations, as implemented in the VASP code (Kresse & Furthmüller, 1996a, b). We employed the generalized gradient approximation (GGA) with the PBE exchange correlation functional (Perdew *et al.*, 1996). In order to improve the description of the highly localised Co 3$d$ orbitals, we employed the so-called GGA+U correction scheme, where we have used a Hubbard parameter $U_{\text{eff}}$ = 6.1 eV, which is the value found for Co 3$d$ by Wdowik & Parlinski (2007) to reproduce the experimental band gap of cobalt (II) oxide (CoO). All calculations were performed allowing spin polarization, as the Co(II) cations formally have the electronic configuration 3$d^7$. We tested both low-spin and high-spin configurations with different magnetic orderings, and found that the Co(II) ions always prefer to be in high-spin configurations (3 unpaired electrons or $S$=3/2), with the magnetic moments being weakly coupled (energy differences between ferromagnetic and antiferromagnetic configurations will be discussed below). The interaction between the valence electrons and the core was described using the projected augmented wave (PAW) method (Blöchl, 1994) in the implementation of Kresse & Joubert (1999). The core levels up to 3$s$ in Ca, 3$p$ in Co, and 1$s$ in C and in O were kept frozen in their atomic reference states. The number of plane waves in VASP is controlled by a cutoff energy, in our case 520 eV, which is 30% higher than the standard value for the PAW potentials employed. For reciprocal space integrations we used a Γ-centred $k$-point mesh of 8, 3 and 2 divisions along the short, medium and long axes of the structures (the corresponding lengths are similar for the malachite and rosasite structures). We checked that these settings of cutoff energy and $k$-point grids lead to total energies converged within 1 meV per formula unit (the convergence in relative energies is likely to be even better). Each structure was fully relaxed (both cell parameters and ion coordinates) to the equilibrium geometry using a conjugate gradients algorithm until the forces on the atoms were all less than 0.01 eV/Å.

## 2.2. Sample preparation and electron microscopy imaging

We synthesised the cobalt hydroxide carbonate using a hydrothermal method to ensure complete crystallization. A 0.05 M aqueous solution of $CoCl_2 \cdot 6H_2O$ was mixed with the same volume of a 0.05 M aqueous solution of $Na_2CO_3$. The mixing was reached in a jacketed glass reactor equipped with an entry for the thermocouple in order to regulate the temperature. The final solution was kept at 65°C in constant stirring for 6 days. After the reaction time, the aqueous solution was cooled at room temperature and then filtered using a 0.45 millipore paper filter. The solid was dried also at room temperature and then powdered in an agate mortar. Although sample preparation at higher temperature could in principle lead to better crystallinity, this is complicated by the formation of $Co_3O_4$. For example, a synthesis attempt increasing the temperature from 65°C to 130°C, for 1 day, failed to produce cobalt hydroxide carbonate and led instead to $Co_3O_4$, as confirmed by Raman analysis.

Scanning electron microscopy (SEM) and transmission electron microscopy (TEM) images were taken in a JEOL 6610LV and a JEOL JEM-2100F microscope, respectively. Each instrument is equipped with an Energy Dispersive X-ray micro analysis system supplied with a silicon drift detector.

## 2.3. X-ray diffraction measurements

Powder X-ray diffraction measurements were made using a Stoe Stadi-P powder diffractometer equipped with a Mo X-ray anode (set to 50 kV, 40 mA), a Ge(111) monochromator providing Mo K$α_1$ (incident wavelength λ = 0.7903 Å), a reduced axial-divergence collimator, and a Mythen 1K detector. Mo X-ray radiation was used instead of the more common Cu X-ray radiation to avoid fluorescence from Co in the sample. The sample was mounted in a 0.5 mm X-ray glass capillary. Diffraction patterns were measured from 1 to 50° in 2θ with a detector step of 0.2° at 120 s per step with the data binned in 0.015° in 2θ. This scan was repeated 5 times to improve the statistical quality of the diffraction patterns and the data totalled.

## 3. Results and discussion

Our DFT calculations started from structures based on experimental data on rosasite (Perchiazzi, 2006), malachite (Susse, 1967) and kolwezite (Deliens & Piret, 1980) minerals, substituting the metal atoms in the original minerals by cobalt. We also used an orthorhombic structure based on the parameters given by (Girgsdies & Behrens, 2012) as a starting point. Upon relaxation, both the kolwezite and the orthorhombic structures converged to the same structure as malachite, while the rosasite converged to a distinct structure. In the language of potential energy landscapes, we can say that the malachite and the rosasite structures are two different local minima, whereas the kolwezite and orthorhombic structures are both within the basin of the malachite minimum. The distinctiveness of the malachite and rosasite structures is clear from the observation that in the former the monoclinic angle is between the short and medium cell vectors, while in the latter it is between the short and long cell vectors. In what follows we deal only with the malachite and rosasite structures, as the other two are unstable.

In order to achieve a fair comparison between the energies of the malachite and rosasite structures, we chose the crystallographic axes for the latter in a way that is different from the setting used originally by Perchiazzi (2006) for the rosasite mineral ($Cu_{1.20}Zn_{0.80}CO_3(OH)_2$) as well as in our previous work on $Co_2CO_3(OH)_2$ (González-López *et al.*, 2016). As can be seen in Figure 1, the monoclinic angle in the rosasite structure can be chosen in different ways, depending on the unit cell definition, and we have simply used the one that gives a value closer to 90° upon relaxation (green cell in figure), since that leads to maximum similarity with the malachite structure.

We have assessed the relative stabilities of rosasite- and malachite-like structures in ferromagnetic and antiferromagnetic configurations for each structure. The Co cations are directly connected by oxygen anions along both the *a* and *c* directions (with reference to the malachite unit cell axes), allowing for superexchange coupling, but are separated by the carbonate species along the *b* direction, leading to an effectively two-dimensional (even if geometrically not flat) network of coupled magnetic centres. Due to the periodicity of the simulation cell, we can enforce antiferromagnetic alternation of the magnetic moments along the *a* direction but not along the *c* direction (in which neighbouring ions are periodic images of one another). Creating a supercell along the *c* direction would allow us to explore different antiferromagnetic configurations, but we have observed that the relative energy of the malachite-like and rosasite-like structure is almost independent of the magnetic configuration, so the consideration of larger supercells is not necessary for the purpose of this study. Table 1 shows that for both structures, the antiferromagnetic configuration is more stable by ~17 meV per formula unit. The rosasite-like and the malachite-like structures are practically degenerate in energy, with a calculated energy difference (~0.05 meV per formula unit) that is too small to be meaningful, considering the general precision of DFT simulations.

We therefore turn to experimental measurements in order to compare (refined) Rietveld models based on the DFT structures with the PXRD patterns. Our cobalt hydroxide carbonate sample obtained at 65 °C is shown in the electron microscopy images in Figure 2. Both the scanning microscopy image (Figure 2a) and the transmission electron microscopy image show well-formed nanocrystals, which exhibit a clear "plate" morphology, in agreement with previous reports (Wang *et al.*, 2009, Zhang *et al.*, 2013).

Figure 3 shows the experimental PXRD diffraction pattern of the sample. Using the DFT-generated malachite and rosasite structures within the Rietveld refinement program Rietica (vers. 1.77), peak position and shape parameters were refined by least-squares fits to the PXRD data with atomic coordinates kept fixed to the DFT values. The calculated pattern for the malachite model is seen in green in Figure 3a and for the rosasite model in red in Figure 3b. Intensity difference plots for both models are shown in Figure 3c. The results show that rosasite-type model gives the best fit to the experimental diffraction data ($R_{wp}$ = 12.9% compared 32.6% for the fit with the malachite model). However, there are still systematic differences in peak intensities between the PXRD data and the rosasite-based Rietveld model, which cannot be resolved by refinement and therefore can be ascribed to the model itself. The refinement of individual atomic coordinates does not result in a significant improvement in the fit to the PXRD data: the $R_{wp}$ can be only slightly reduced by full refinement (from 12.9% to 12.6%), but the resulting coordinates are no more reliable than the DFT ones, since the refinement simply attempts to correct for the peak intensities that cannot be fully described by the rosasite model. Tables 2 and 3 show the DFT-calculated and the Rietveld-refined cell parameters, as well as the atomic coordinates from DFT, for the rosasite and malachite models, respectively.

It is interesting to note here that Perchiazzi & Merlino (2006), in their study of the related compound $Mg_2CO_3(OH)_2$, discussed its possible non-stoichiometry in the form of metal cation vacancies. We have also considered here the refinement of the $Co_2CO_3(OH)_2$ structure varying the site occupancies for both Co(1) and Co(2) positions in the rosasite structure. For Co(2), the site occupation number stays at around 100% and the *R*-factor does not improve. Interestingly, for Co(1), the occupancy drops to around 87% with a 1% improvement in $R_{wp}$. However, $R_{wp}$ is still relatively high at 11.9% because the most intense peak is still poorly fitted by the model. We therefore believe that this result, although interesting enough to be reported, should not be taken as a strong suggestion of the presence of Co vacancies in the cobalt hydroxide carbonate. Given the limitations of the rosasite model, anything that slightly improves the intensity of the most intense peak will reduce $R_{wp}$, so the fractional occupancy may be simply an artefact of the fit. The potential presence of cation vacancies in this compound requires further investigation in future work.

Finally, we discuss possible reasons to why neither the rosasite nor the malachite model gives a completely satisfactory fitting of the PXRD data. A possible explanation, consistent with the small DFT energy difference between the two structures, is that both phases coexist in the sample. However, a two-phase Rietveld refinement does not significantly improve the fit (as measured by $R_{wp}$ and by visual appearance). The refined scale factors from the two-phase model show that the amount of malachite phase present, if any, is insignificant. A closer look at both structures offers a more interesting possible explanation. Figure 4 shows the two structures in a plane perpendicular to the (malachite) *a* axis (the rosasite axes have been redefined again here to show the analogy with malachite). They can be seen as structures made up of identical layers but with different stacking sequences. The relative lateral shifts from one layer to the next are always the same in each structure, involving a ¼ shift along the malachite *c* axis. But while in malachite consecutive shifts are in opposite directions, leading to an ABAB sequence, in the rosasite the shifts are always in the same direction, leading to an ABCD sequence. Therefore the two structures can be considered as polytypes. The fact that not only the layer structure but also the local geometry of the interface is the same for both structure explains their very similar energies: the only difference between the two structures is in the interaction between next-nearest layers. Our results therefore suggest that actual samples might exhibit stacking disorder, with random relative directions of consecutive shifts, instead of the two well-ordered shifting patterns represented by the malachite- and rosasite-like structures. This interesting possibility requires further theoretical and experimental investigation. For the moment, the rosasite-like model reported here is the best available model for the $Co_2CO_3(OH)_3$ structure.

**Acknowledgements**   This research was supported by the Ministry of the Spanish Ministry of Economy and Competitiveness (MINECO) under Contracts CGL2010-20134-CO2-02 and CGL2013-47988-C2-2-P. We made use of ARCHER, the UK's national high-performance computing service, via RGC's membership of the UK's HPC Materials Chemistry Consortium, which is funded by EPSRC (EP/L000202).

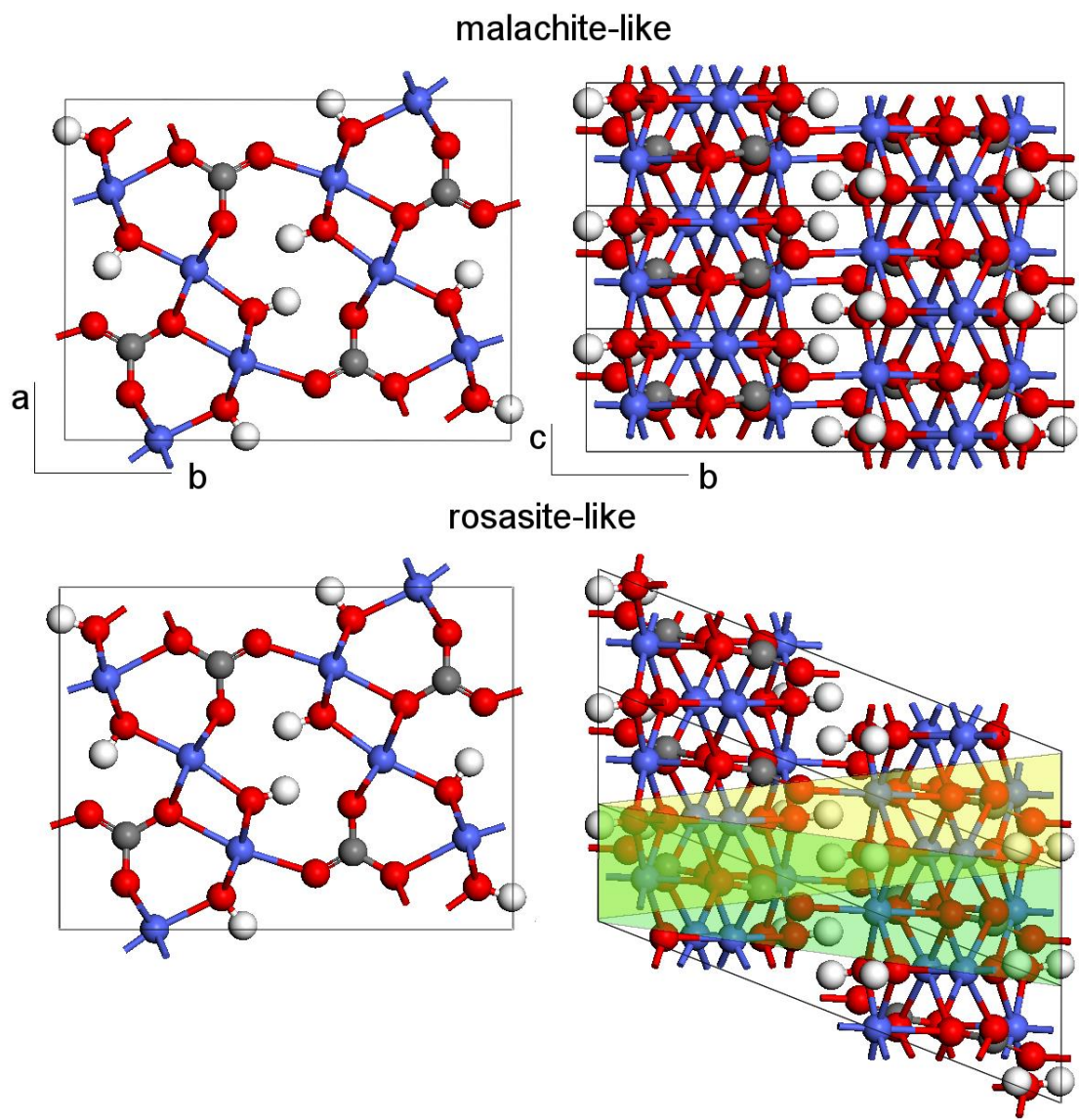

**Figure 1** Malachite-like (top) and rosasite-like (below) crystal structures of $Co_2CO_3(OH)_2$ as obtained from DFT calculations. The rosasite-like structure is displayed with the atomic positions shifted in a way that maximises the coincidence with the malachite structure and does not follow the values listed in Table 2. Colour shading is used to represent alternative cells with different values of the monoclinic angle. The green-shaded cell was used for the DFT calculations. Colour code: blue = cobalt; grey = carbon; red = oxygen; white = hydrogen.

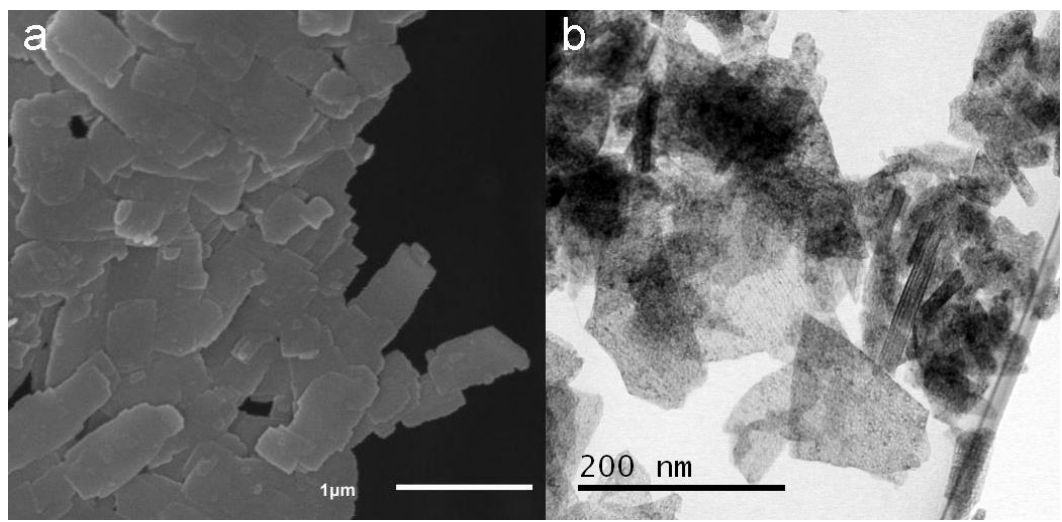

**Figure 2** Scanning electron microscopy (a) and transmission electron microscopy (b) images of $Co_2CO_3(OH)_2$.

To appear in Acta Crystallographica B

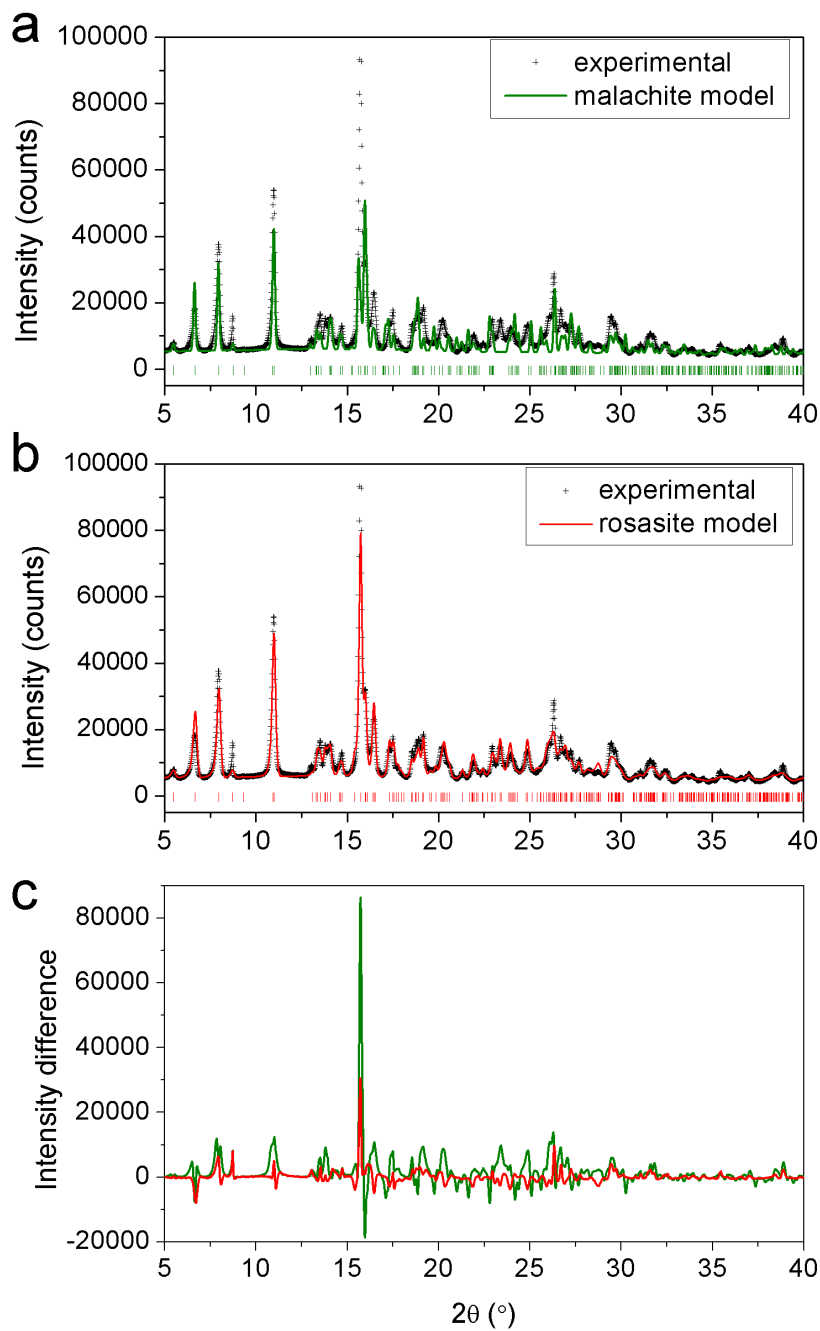

**Figure 3** Experimental X-Ray diffraction pattern ("+" symbol) compared with a) malachite-like and b) rosasite-like (green and red lines respectively) Rietveld refinement curves (atomic positions fixed to DFT values). c) Difference between experimental and refined intensities for both models.

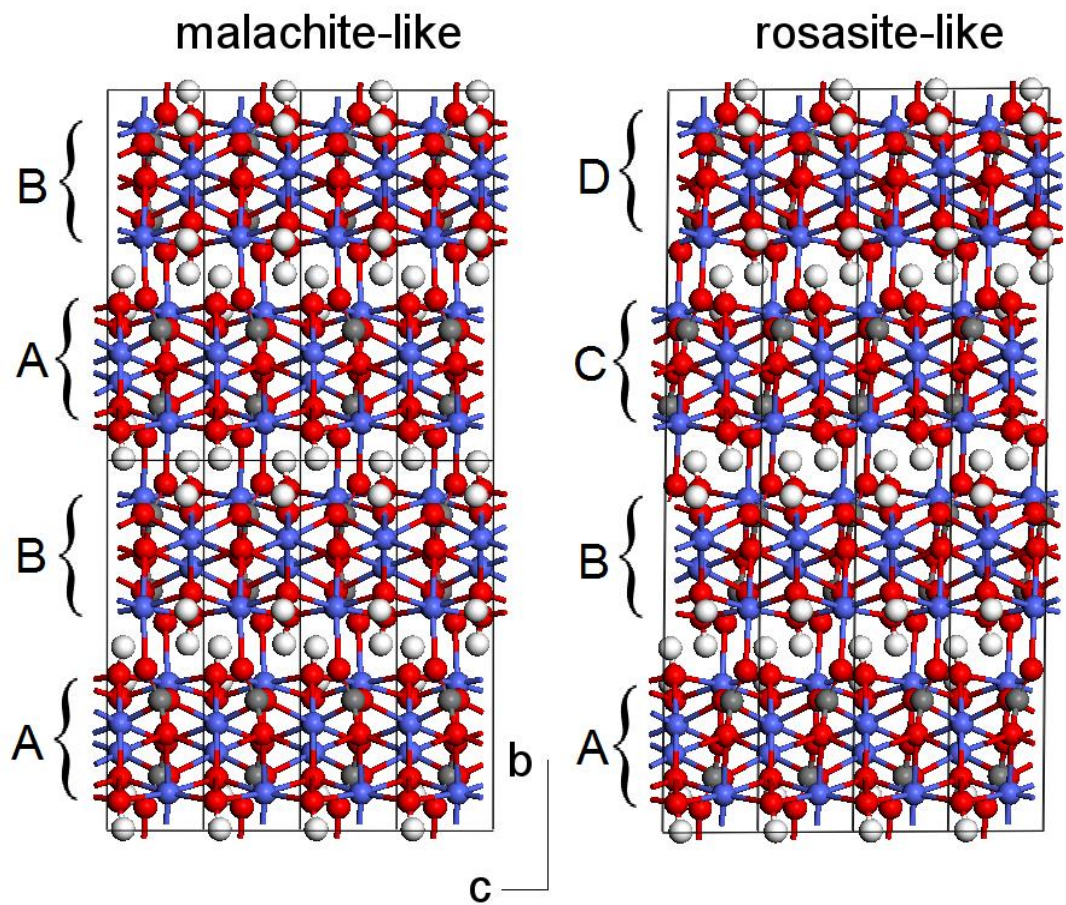

**Figure 4** Malachite-like and rosasite-like structures of $Co_2CO_3(OH)_2$ seen as two different stacking sequences of the same two-dimensional motif. The rosasite-like structure is shown using a redefined supercell lattice for better comparison with the malachite-like structure. Colour code as in Figure 1.

**Table 1** Relative DFT energies for the malachite-like and rosasite-like structures of $Co_2CO_3(OH)_2$ in the ferromagnetic (FM) and antiferromagnetic (AFM) configurations described in the main text.

| Structure | E (meV/fu) | |
|---|---|---|
| | AFM | FM |
| Malachite | 0 | 16.92 |
| Rosasite | 0.04 | 16.98 |

**Table 2** Cell parameters and atom coordinates of $Co_2CO_3(OH)_2$ in a rosasite-like structure, as obtained from DFT calculations (Rietveld-refined values of cell parameters are given within the square brackets).

| | Rosasite-like structure | | |
|---|---|---|---|
| Space group | $P112_1/n$ | | |
| $a$ / Å | 3.174 [3.1408(4)] | | |
| $b$ / Å | 12.374 [12.2914(17)] | | |
| $c$ / Å | 9.413 [9.3311(16)] | | |
| $\gamma$ / degrees | 82.82 (82.30) | | |
| Coordinates | $x$ | $y$ | $z$ |
| Co1 | 0.77660 | 0.71075 | 0.49778 |
| Co2 | 0.18314 | 0.89784 | 0.26841 |
| C | 0.38881 | 0.64742 | 0.22817 |
| O1 | 0.30694 | 0.64639 | 0.36513 |
| O2 | 0.28751 | 0.73926 | 0.15774 |
| O3 | 0.57404 | 0.56386 | 0.16515 |
| O4 | 0.70019 | 0.85789 | 0.40510 |
| O5 | 0.67413 | 0.91997 | 0.12379 |
| H1 | 0.31773 | 0.00498 | 0.90820 |
| H2 | 0.27813 | 0.09536 | 0.51076 |

**Table 3** Cell parameters and atom coordinates of $Co_2CO_3(OH)_2$ in a malachite-like structure, as obtained from DFT calculations (Rietveld-refined values of cell parameters are given within the square brackets; however, note that the quality of the fit with this model is poor – see text).

|  | *Malachite-like structure* | | |
|---|---|---|---|
| Space group | $P12_1/a1$ | | |
| $a$ / Å | 9.425 [9.307] | | |
| $b$ / Å | 12.261 [12.224] | | |
| $c$ / Å | 3.174 [3.135] | | |
| $\beta$ / degrees | 91.12 [90.49] | | |
| Coordinates | $x$ | $y$ | $z$ |
| Co1 | 0.00262 | 0.28894 | 0.86602 |
| Co2 | 0.73213 | 0.39792 | 0.3694 |
| C | 0.77217 | 0.14755 | 0.45237 |
| O1 | 0.63493 | 0.1467 | 0.36907 |
| O2 | 0.84296 | 0.2389 | 0.38999 |
| O3 | 0.83503 | 0.06414 | 0.60326 |
| O4 | 0.59572 | 0.3581 | 0.86532 |
| O5 | 0.87648 | 0.42001 | 0.8779 |
| H1 | 0.51153 | 0.40484 | 0.81598 |
| H2 | 0.90773 | 0.49536 | 0.83501 |